\documentstyle[epsf]{article}

\newcommand{\be}{\begin{equation}}
\newcommand{\ee}{\end{equation}}

\begin{document}

\begin{flushright}
Liverpool Preprint: LTH 566\\
 \end{flushright}
  
\vspace{5mm}
\begin{center}
{\LARGE \bf Hadronic decay of a vector  meson from the lattice}\\[10mm] 
{\large\it UKQCD Collaboration}\\[3mm]
 
 {\bf C. McNeile and    C.~Michael \\
Theoretical Physics Division, Dept. of Mathematical Sciences, 
          University of Liverpool, Liverpool L69 3BX, UK 
 }\\[2mm]

\end{center}

\begin{abstract}

 We explore the decay of a vector meson to two pseudoscalar mesons  on
the lattice with $N_f=2$ flavours of sea quark. Although we are working
with quark masses that do not allow a physical decay, we show how  the
transition rate can be evaluated from the  amplitude for $\rho
\to \pi \pi$  and from the annihilation component of $\pi \pi \to \pi
\pi$. We explore  the decay amplitude for two different pion momenta and
find consistent results. The coupling strength we find is in agreement
with experiment.  We also find evidence for a shift in the $\rho$ mass
caused  by mixing with two pion states.

\end{abstract}

\section{Introduction}

 The study of  hadronic decays  using lattice techniques has  long been
known to be feasible in principle~\cite{cmdecay,lu91}. Pioneering
attempts were made to study  glueball decay~\cite{wein} and $\rho$ meson
decay~\cite{degrand} in quenched QCD where no actual decay takes place.
This underlines the approach: one can study  the mixing of hadronic
states on a lattice, provided that the energies  of the states are close
(see ref~\cite{cmcmscalar,hdecay} for quantitative analysis). Indeed 
string breaking has been explored this way~\cite{cmadj,cmpp}  as has 
scalar meson mixing~\cite{cmcmscalar} and hybrid meson
decay~\cite{hdecay}. The common feature  of these examples is that there
is little experimental knowledge of the  relevant transition matrix
elements. Here we rectify this by studying  $\rho$ meson decay. Another
motivation is that non-leptonic weak decays ($K \to \pi \pi$ especially)
are very important to understand from non-perturbative QCD and a study 
of simpler purely hadronic decays will be a useful step in this
direction.

\begin{figure}[th]
\epsfxsize=10cm\epsfbox{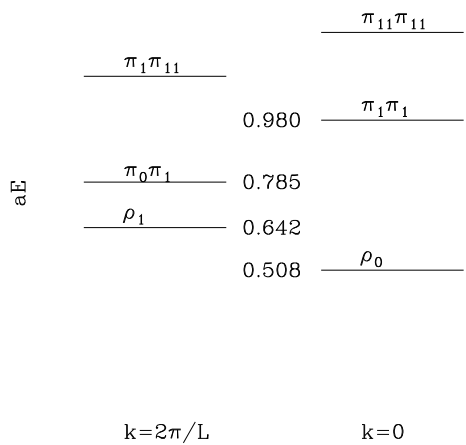}
\vspace{-1cm}
 \caption{ The energy spectrum (in lattice units with $a \approx 0.11$
fm) on a lattice of the $\rho$ meson and  two pion  states considered
here. Here $k$ is the overall  momentum  and the  suffixes are the
components (in units of $2\pi/L)$ of momentum of  the mesons. The energy
values  come from UKQCD fit~\cite{ukqcd} central values for zero
momentum states.
 }
 \label{fig.rpp}
\end{figure}

 On a lattice many features are different from in the real world. The
most  significant for our purposes is that periodic spatial boundary
conditions are  imposed. Note that the continuum limit in such a finite
volume is defined and, if the  spatial  extent is  $L$,  the momentum is
discrete in units of $2 \pi/L$. This implies that the two  particle
spectrum is discrete. For decay of a vector meson in the centre of mass,
 in a P-wave (actually the $T_1^{--}$ representation of the cubic 
rotation group), the decay momentum must be non-zero and hence the
lightest $\pi \pi$ state will be with momentum 1 and -1 in these units. 
 For the lattices we will use,  the energy levels neglecting
interactions are  illustrated in fig.~\ref{fig.rpp}. As has been  noted
before~\cite{degrand}, a closer match in energy can be achieved by
considering the decay of a moving vector meson, see also
fig.~\ref{fig.rpp},  since this effectively allows a relative momentum
in the centre of mass of  $\pi/L$. Note that in neither case is the
$\rho$ meson unstable.

\begin{figure}[th]
\epsfxsize=9cm\epsfbox{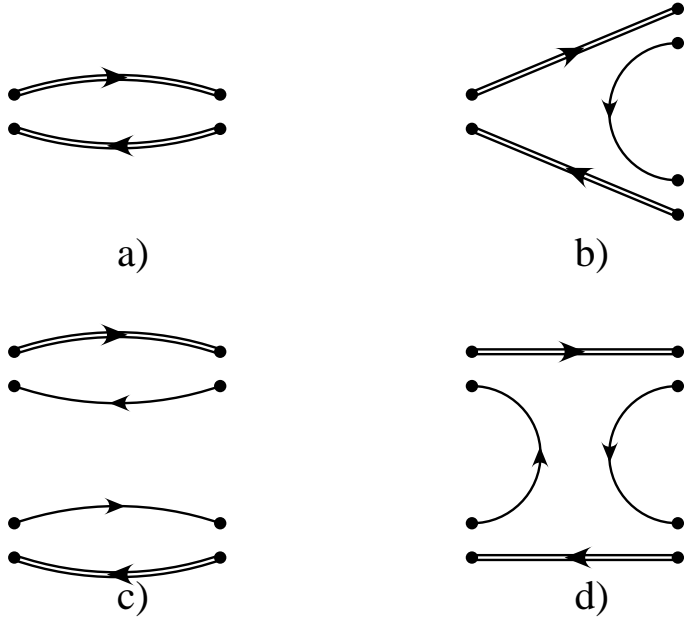}
 \caption{ The quark diagrams studied here, where (a) is $\rho$
propagation,  (b) is $ \rho \to \pi \pi$, (c) and (d) are the direct and
`box' components of  $\pi \pi \to \pi \pi$. The valence quarks are shown
by a  double line and the  quark-antiquark pair created in the decay  by
a single line, although in this work  both quarks are  taken as having
the mass of the sea quark (approximately of strange quark mass). As
described in the text, we take the P-wave component by  antisymmetrizing
on exchange of the final state pions. Note that the crossed  quark
diagram ( $\overline{\underline{\rm X}}$ )does not contribute.
 }
 \label{fig.quark}
\end{figure}

 As an illustration, consider the transfer matrix with two states, 
a $\rho$ meson with energy $m-\Delta/2$ and a $\pi \pi $ state 
with energy $m+\Delta/2$ with a transition amplitude  $x =
\langle \rho | \pi \pi \rangle$: 

\be
 e^{-ma}
 \left(
 \begin{array} {cc}
 e^{-a\Delta/2} &  ax \\
 ax & e^{a\Delta/2} 
 \end{array}
 \right)
\ee

 \noindent  This transfer matrix  has eigenvalues $\lambda= e^{-Ea}$
with 
 \be
    E \approx m \pm  (\Delta^2/4 +x^2)^{1/2}
 \label{eq.eig}
 \ee
 \noindent  which is larger  than the unmixed splitting by a shift of
$\epsilon=(\Delta^2/4 + x^2)^{1/2}-\Delta/2$  up (assuming $\Delta > 0$) of the
two pion state and down for the $\rho$ state.

We  expect the lightest state with the relevant quantum numbers (isospin
1,  $T_1^{--}$ representation) to be dominantly created by a local (or
fuzzed)  quark antiquark operator, as conventionally used to study the
$\rho$ meson. Likewise, the lightest two pion state is expected to be
dominantly created by  a lattice operator made of two pion operators.
These statements are qualitative,  and L\"uscher has
emphasised~\cite{lu91} that a quantitative description can be based  on
a careful measurement of the two-particle energy levels (using in
principle any lattice operators whatsoever) for different sizes $L$. 
 To determine this energy shift accurately is a challenging  task. As
well  as attempting to determine it directly, we measure the mixing
$x=\langle \rho | \pi \pi \rangle$ which allows us to estimate the
energy shift from eq.~\ref{eq.eig}, as well as to estimate 
the transition amplitude directly.

 The essence of the determination of this transition amplitude is that
when the energies of the  two hadronic states ($\rho$ meson and $\pi
\pi$ system with the same quantum numbers) are close (specifically when
$t \Delta  \ll 5$) then the exponentials in $t$ corresponding to the two
energy states conspire to give a linear dependence (see ref.\cite{cmcmscalar}
for a discussion). The transition measured on a lattice then gives
access~\cite{cmcmscalar,hdecay} to the  required matrix element since 
for large $t$ (specifically $(E'-E)t \gg 1$ with $E'-E$ the energy gap
to  the first excited state),
 \be
  { \langle \rho(0) | \pi(t) \pi(t) \rangle  \over
 \langle \rho(0) | \rho(t) \rangle^{1/2} 
\langle \pi(0) \pi(0) | \pi(t) \pi(t) \rangle^{1/2}  }
 = xt + {\rm const}
 \label{eq.xt}
 \ee
 provided that the transition rate is not too large, namely $xt \ll 1$.

 A further cross check~\cite{cmcmscalar,hdecay} is also possible from
the box diagram (see fig.~\ref{fig.quark}d) under similar conditions,
since
 \be
  { \langle \pi(0) \pi(0) | \pi(t) \pi(t) \rangle_{\rm box}  \over
\langle \pi(0) \pi(0) | \pi(t) \pi(t) \rangle_{\rm direct}  }
 = {1 \over 2} x^2 t^2 + {\cal O}(t)
 \label{eqxx}
 \ee
 Here the notation $\pi(t),\ \rho(t)$, etc refers to creating a state
with those  quantum numbers from the vacuum at that time on the lattice.
The denominators are to normalise the states to unity. To relate  the
matrix element $x$ to the usual continuum large volume formalism, one
has to  relate this unit normalisation condition to the usual
relativistic one - see ref~\cite{lu91,ll,testa} for a full discussion. One
simple way to  do this is by considering the formula for the decay
width, though we must emphasise  that no actual decay takes place in our
case because of the unrealistic quark masses.
 Then first order perturbation theory (Fermi's Golden Rule) implies  a
transition rate $\Gamma=2 \pi \langle x^2 \rangle \rho(E)$ where the
angle brackets indicate that an average over spatial directions will be
needed. For a decay from the centre of mass with relative momentum $k$,
the density of states  $ \rho(E)= L^3 k E /(8 \pi^2)$.

 We  use the mixing we establish on the  lattice to evaluate the mass
shift $\epsilon$ of the two-particle state. This mass  shift turns out
to be too small to measure accurately by a direct determination for the
two-pion state, but  we  are able determine this shift from a study of
the $\rho$ mass.  From our estimate of the energy shift, we are able to
use L\"uscher's formalism  to determine the $\rho$ decay parameters,
obtaining results in agreement  with the method described above. 

 Here we present our preliminary study of this problem. We establish
methods that give  a good signal on a lattice and discuss the
cross-checks that can be made. Our analysis is restricted to 
mesons made of quarks of mass close to the strange quark mass. We 
discuss the phenomenological implications. Overall we find that  lattice
study of hadronic transitions is feasible and we find  good agreement
between our determinations and the expectations from experiment.
 The vector meson transition to two pseudoscalar mesons is actually 
rather strong and our methods will apply even better to weaker 
transitions, such as hybrid meson decay~\cite{hdecay}.

\section{Lattice results}

 We use the UKQCD data set with $N_f=2$ flavours of sea quarks with a 
 NP clover fermionic action  and Wilson glue at  $\beta=5.2$ and
$C_{SW}=2.0171$,  $\kappa_{\rm sea}=\kappa_{\rm valence}=0.1355$, volume
$16^3\times 32$ and configurations separated by 40
trajectories~\cite{ukqcd}. This corresponds~\cite{ukqcd} to a quark mass
 for which $m_{\pi}/m_{\rho} = 0.578^{+13}_{-19}$ which is 
approximately the strange quark mass, with a lattice spacing
$a=0.110(4)$ fm.

 We define  $\rho$ and $\pi \pi$ states as being in a given $T_1^{--}$
representation of the cubic group in the centre of mass. We consider the
$\rho$ meson state as having polarisation in a given direction (eg 
using $\bar{q} \gamma_z q$ to create it). We shall consider $\rho^+ \to
\pi^+ \pi^0$ so that the pions are distinguishable and this has 
contributions from two triangle quark diagrams (see
fig.~\ref{fig.quark}(b)) with $u\bar{u}$ and $d\bar{d}$ creation
respectively: each with factors of $ 1/\sqrt{2}$ from the $\pi^0$. We
normalise the P-wave $\pi \pi$  state with relative momentum $k$ in the
centre of mass as $[\pi^+(k) \pi^0(-k)-\pi^+(\bar{k})
\pi^0(-\bar{k})]/\sqrt{2}$ where $\bar{k}$ has the $z$-component of $k$
reversed. In the cases we  will consider, $k$ only has a $z$-component,
so  $\bar{k}=-k$. In order to have a smaller relative momentum, we also
consider transitions from a $\rho$ with nonzero momentum (where we take
the polarisation along the momentum direction). In that case we define 
$x$ as $\langle \rho_1 | \pi_1 \pi_0 - \pi_0 \pi_1 \rangle/\sqrt{2}$,
also the $z$-axis is now privileged  and we study the $A_2^-$
representation of the symmetry group $D_{4h}$.

For the three and four-point correlators, we use a stochastic method
to evaluate them  from every space-time point on each lattice. Because
of this  volume averaging, we are able to get results from only 10
lattice configurations.
 The stochastic sources are Gaussian on one timeslice only~\cite{fm} and
extended  propagator techniques are used to evaluate 3 and 4-point
correlators. This study of the three and four-point correlators
involved 2880 inversions which is comparable to the number of inversions
needed to evaluate propagators from one space-time point on 208
configurations (as used for the two-point correlators~\cite{ukqcd}).

 We find that the $\pi \pi \to \pi \pi$ P-wave amplitude is  dominated
by the  product of two two-point pion correlators (see
fig.~\ref{fig.quark}c) which is  accurately known,  whereas the
correlations between the pions  and the box contributions are  more
noisy. Thus we choose to normalise by the two-point pion correlators.
Moreover, in order to suppress contributions from excited states we 
normalise by the ground state contributions to the ($\rho$ and $\pi$) 
two-point correlators as determined by 2-state fits to the UKQCD
correlator data~\cite{ukqcd} (with local and fuzzed sources and sinks)
for a $t$-range of 4-12.

\begin{figure}[th]
 \centerline{\epsfxsize=8cm\epsfbox{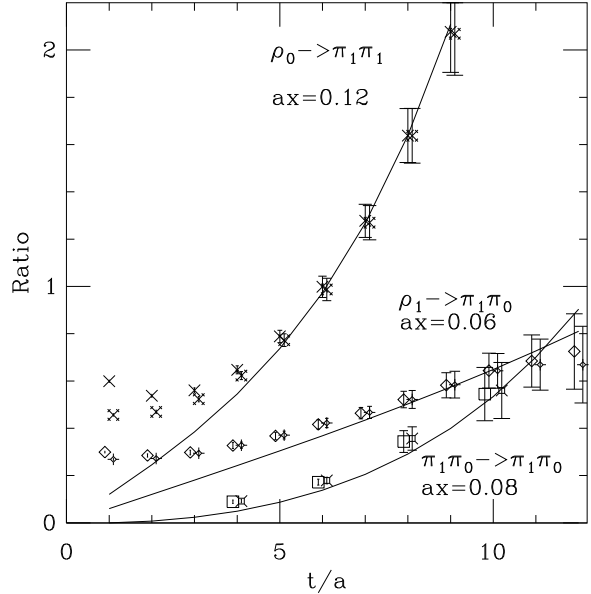}}
 \caption{ Normalised three and four particle correlators  versus $t/a$.
 For the three particle ($\rho \to \pi \pi$) case,  the crosses(fancy
crosses) are for relative  momentum $2\pi/L$ with local(fuzzed)
operators while the diamonds(small diamonds) are for $ \pi /L$ with 
local(fuzzed) $\rho$ operators. The curves are with transition
amplitudes $ax=0.12$ (upper) and $ax=0.06$ (lower) as described in the
text. For the  P-wave annihilation component of the four particle pion
correlator (the `box diagrams') with pion  relative momentum
$\pi /L$, the results are  shown by squares(fancy squares) for local
pion operators (one pion fuzzed). The model curve is shown here for
$ax=0.08$.
 }
 \label{fig.xt} 
 \end{figure}

\begin{figure}[th]
 \centerline{\epsfxsize=8cm\epsfbox{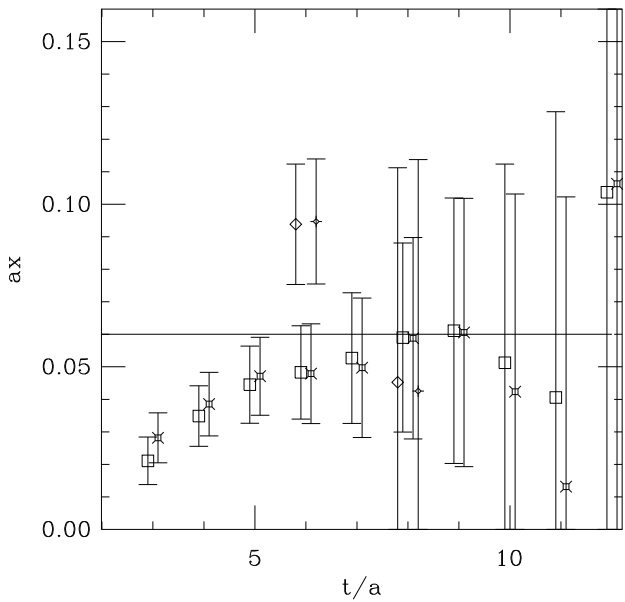}}
 \caption{ The matrix element $ax$  versus $t/a$ for the transition
$\rho_1 \to \pi_1 \pi_0$.  These  results come from  finite differences
to remove excited state contaminations in the  data  of
fig.~\ref{fig.xt}. The squares and fancy-squares are from the
three-point analysis (with  local and fuzzy $\rho$ operator
respectively) with a finite difference  taken over two $t$ intervals.
The diamond (small diamond) is from the  Box diagram with all local 
$\pi$ operators (one fuzzed)  evaluated by subtracting  linear and
constant terms using $t/a=4-8$, and $6-10$.
 }
 \label{fig.x0}
\end{figure}

We present our results for three-point correlators  from eq.~\ref{eq.xt}
in fig.~\ref{fig.xt}. Note that the relevant ratio is quite  large,
becoming comparable to unity at larger $t$. The requirements of our
method should be satisfied for $3 \ll t \ll 12$ for  $\rho_1 \to \pi_1
\pi_0$, since we require $xt \ll 1$, $ \Delta t \ll 5$ where $a\Delta =
0.14$ and $ (E'-E)t \gg 1$ where $E'-E \approx 0.3$.  The results in
this case have quite small errors and do show well  the required linear
dependence on $t$ in this region. The curve is from the two-state
transfer matrix model with $ax=0.06$. The departure at small $t$ is
presumably due  to excited state contributions to the three-point
correlator. These  cannot produce a linear dependence on $t$, however,
although they can contribute a constant even at large $t$. Since such
constant contributions from excited states are allowed in principle,   
we show  in  fig.~\ref{fig.x0} the slope of the correlator ratio of
eq.~\ref{eq.xt}. This  is consistent with  constant at $ax=0.06$ in this
region, for both local and  non-local $\rho$ sources.

 As emphasised before~\cite{cmcmscalar}, a cross check is available from
the box  diagram (eq.~\ref{eqxx}). Note that we  need to evaluate this
for both  the momentum direct ($10 \to 10$) and crossed ($10 \to 01$).
These contributions  are computationally difficult  to measure  and we
have chosen $t$ values of 4, 6, 8 and 10 here, as illustrated in
fig.~\ref{fig.xt},  again normalising by the ground state two-point
correlators, where a comparison is made with the two-state transfer
matrix model  with $ax=0.08$. Since excited state contributions can
produce both constant  and linear terms for this quantity, we form the
linear combination of  three adjacent $t$-values that eliminates them
and so determine $x$ by  comparing with the two-state transfer matrix
model. The resulting values are shown in fig.~\ref{fig.x0} where they
are seen to be compatible with those from the  three particle analysis.

 Thus we conclude that we do have a consistent lattice determination of 
$ax= 0.06^{+2}_{-1}$ for this momentum combination. 
 From this determination of the transition amplitude, we can determine 
an energy shift,  as given  by eq.~\ref{eq.eig}, assuming that only the
two nearest levels mix. This yields a shift of 
$a\epsilon=0.022^{+17}_{-7}$  - up for the  $\pi_1 \pi_0$ state and down
for the $\rho_1$ state.

\begin{figure}[th]
 \centerline{\epsfxsize=8cm\epsfbox{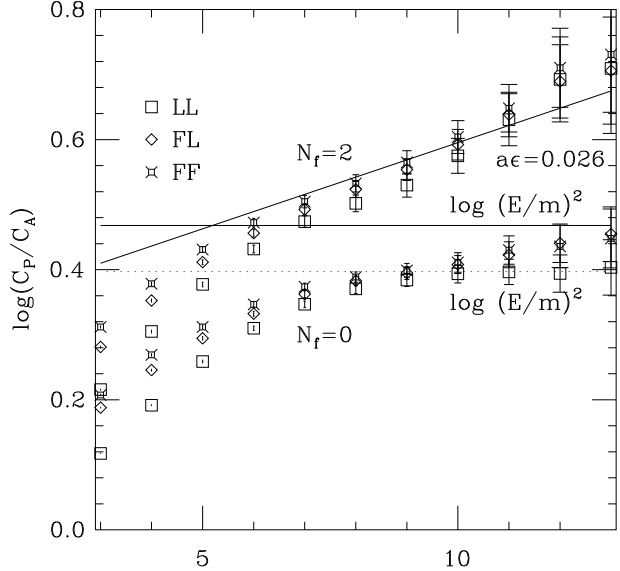}}
 \vspace{-0.2cm}
 \caption{ The ratio of two-point $\rho$ correlators with momentum
$2\pi/L$ and  polarisation parallel (P) or perpendicular (A) to the
momentum. These are for sources/sinks as shown from UKQCD~\cite{ukqcd} 
configurations with $N_f=2$ (upper) and quenched (lower with similar
quark masses at $\beta=5.93$  and $\kappa=0.1339$). If there were no
mixing between these $\rho$ states and  the nearby $\pi_0 \pi_1$ state,
the ratio should be $(E/m)^2$ (shown as a line) for the  ground state
and nearer to 1.0 for heavier excited states. Thus we would expect the
ratio to reach a plateau at $t> 6$ where the ground state dominates.
Instead for the dynamical quark data, we see a significantly higher
ratio and  the line shows the slope obtained from the mass splitting
$\epsilon$ we find in fitting. For quenched lattices  the ratio just
rises  to $(E/m)^2$ (shown dotted) at larger $t$ but does not cross it.
 }
 \label{fig.rhoratall}
\end{figure}

 For the $\rho$ meson, a general investigation of  the energy shift due
to  two pion intermediate states is complicated by  the need  to
regulate the sum over pion momenta.
 Here we are interested primarily in  shifts arising from significant
mixing with nearly degenerate  energy levels in a finite volume. One
signature~\cite{milc} of such a shift is that  the cubic invariance will
be broken: the energy of a $\rho$ meson with  momentum $2\pi/L$ will be
different when it is polarised in the  momentum direction or
perpendicular to it.
  We have explored this energy shift directly from the study of  2-point
$\rho$ correlators with momentum $2\pi/L$. When the polarisation is
along the momentum,  there will be mixing with the nearby $\pi_1 \pi_0$
state for dynamical simulations (but not for the quenched case), while 
when the polarisation is perpendicular there will be no such mixing.
From fits to the correlations (using 207 configurations with local and
fuzzed sources at 4 time-points and requiring  the excited  state mass
to be the same for the two fits to the $t$-range 4-12), we  see a
significant mass shift: namely the parallel $\rho$ is 
$a\epsilon=0.026(7) $ lighter than the perpendicular. We illustrate this
in fig.~\ref{fig.rhoratall}. This value is in excellent agreement with
the mass shift we deduced above by mixing arguments.  We measured this
mass shift for other sea quark masses, but have a significant signal
only for the case  ($\kappa=0.1355$) studied here. Note that this
directly observed mass-shift  for the $\rho$ state  should  not be
present in a quenched study and it is not, see fig.~\ref{fig.rhoratall},
so we have here one of the few  observables that directly come from the
sea-quark contributions.

An even more useful result would be a direct determination of  the
energy shift for the $\pi \pi$ state. The optimum way to determine the 
energies of the two close levels  would be by a joint fit (or
variational analysis) of the matrix of  $\rho$ and $\pi \pi$
correlations. For this we need the 4-point $\pi \pi$ correlator, for
which the ratio of contributions to the product of 2-point pion
correlators at $t=8$ is 1.07(4) for the direct term, -0.04(4) for the 
momentum-swapped direct term and 0.35(6) for the box terms. The relative
error on the total 4-point correlator  at $t=8$ is $6\%$. From fitting
the  matrix of correlators for $t$ from 8 to 12 with two states, we 
form the combination orthogonal to the ground state, so selecting the 
first excited (ie mainly $\pi \pi$) state.  From this we evaluate  the
`un-binding energy', the  $\pi \pi$ energy  difference from the sum of
pion energies, finding $a\epsilon=0.02(2)$. This value is in excellent
agreement with our values obtained from the mixing analysis and from the
$\rho$ mass shift.  The error, however, is still extremely large and a
major increase in  computational resource would be needed to determine
this energy shift directly  with sufficient precision.

 For the case when $\rho_0 \to \pi_1 \pi_1$, the energies are further
apart,  the next excited state is closer (see fig.~\ref{fig.rpp}) and
the  expected value of $x$ is twice as big since it increases like the
relative momentum $k$ for a P-wave decay. We estimate that  $4 \ll t \ll
8$ is required in this case. As shown in fig.~\ref{fig.xt}, the values 
of the three-point correlator  obtained for this case are approximately
twice as large, but there  is no longer a region showing linear
behaviour. We show the result from the  two-state transfer matrix model
with the expected larger energy gap ($a\Delta=0.47$ in this case) and
normalised by the unmixed $\pi$ two-point correlators (as we have used
in the figures) and this gives  the curve shown in the figure for
$ax=0.12$. This value is close to twice the value obtained with  half of
the relative momentum above, as expected. In this latter case, however,
we  do not have substantial cross-checks and this is qualitative rather
than quantitative.

 The method we have used to determine $x$ depends on eliminating excited
state contributions that appear  at subleading powers of $t$. It is 
impossible to exclude that a combination of such terms do modify our
results significantly. However, the cross checks that we have available
are  strong: we see consistent values for $x$ from local or fuzzed
sources from 3-point correlators, from the  box contribution to the
4-point correlator  and from the 3-point correlator with larger
momentum release. We also see  consistent evidence for the energy shift
of the $\rho$ meson with different polarisation directions.

\section{Phenomenology}

 We have evaluated the transition $\rho \to \pi \pi$ at an unrealistic 
quark mass (approximately the strange quark mass), in a finite  volume
(of size $16a \approx 1.76 $fm) and with $N_f=2$ flavours of sea quark.
Note that the lattice technique enables the transition to be measured
off the energy-shell.

For the $\rho_1 \to \pi_1 \pi_0$ case, the relative momentum in the
centre of  mass (with boost given by $\gamma=1.156$ to the lattice
frame) corresponds to $ka=\pi/L\gamma$, hence 305 MeV/c  which is  close
to the experimental momentum of 358 MeV/c for $\rho \to \pi \pi$. 

 We have normalised the P-wave $\pi \pi$ state so that  relative
momentum $k$ and $\bar{k}$ are identified which reduces the density  of
states by 0.5.   Since we have evaluated $x$ for  momentum aligned with
the polarisation, there will be an angular average contributing a factor
of 1/3. Hence we expect
 \be
\Gamma=  x^2   L^3 k E /(24 \pi)
 \label{eq.width}
 \ee

 Our lattice determination is most precise for the transition $\rho_1
\to \pi_1 \pi_0$  which is not in the centre of mass. We proceed first
by evaluating the  decay width in the lattice frame. 
We use the expression of eq.~\ref{eq.width} with $k=\pi/L$, the
relative  momentum between the pions in the lattice frame. There are
additional  factors of $\gamma$ from the phase space and from
transforming the  width to the centre of mass frame.  Then, defining a
dimensionless coupling constant directly from the reduced width as 
$\bar{g}^2=\Gamma M E/k^3$ (corresponding to $g^2/(6\pi)$ where $g$
is the conventional definition  of the coupling constant), we find  a
value of $\bar{g}=1.40^{+47}_{-23}$ from $ax=0.06^{+2}_{-1}$, where we
have used our  best estimates of the  centre of mass values $E$ and $k$
as discussed below.

 One can relate the lattice frame to centre of mass frame  by
considering the boost of the two pions to the centre of mass (with
$K=2\pi/L$ and $\gamma = (1-K^2/E^2)^{-1/2}=1.156$). Then in the centre
of mass, the cubic  spatial volume is modified by  $L \to \gamma L$ in
the direction of $K$ (taken as the $z$-axis here). Moreover  in this
extended cube, pion momenta $k=2 \pi n/ \gamma L$ are allowed with 
$n_z$ half-integral, coming  from anti-periodic spatial boundary
conditions in the relative pion $z$-coordinate. Although this boost (by
$\gamma$)  reduces the two pion  system to the centre of mass, the
$\rho$ will not have zero momentum.  This arises  since  energy is not
conserved in the lattice frame, as we measure $t$-directed correlations,
which implies that momentum is not conserved  in the centre of mass.  So
our estimate above is uncertain by factors of $\gamma(\pi
\pi)/\gamma(\rho) = 1.156/1.264$.
 As the energy gap between the $\rho$ and its decay products gets 
smaller, this problem will decrease. The transition $\rho_0 \to \pi_1
\pi_1$  is in the centre of mass, which avoids the above problem, and
$ax=0.12$ corresponds to $\bar{g}=1.57$ which  is consistent, although
systematic errors in this case are large.

 A more rigorous approach is to focus on the two pion energy as in  the
L\"uscher formalism~\cite{lu91}. This has been generalised ~\cite{rumm}
to non-zero  overall momentum by considering the boost to the centre of
mass.  Although we are unable to determine  the energy shift $\epsilon$
directly for the two-pion state, if we assume that two levels only
dominate the mixing, then we can use our results from the $\rho$ meson
energy shift and from the shift deduced through mixing from our value of
$x$.  Taking $a\epsilon=0.26(7)$ yields  an energy shift in the centre
of mass frame of $\epsilon \gamma$ (since  $E_{L}^2=E_{\rm cm}^2+K^2$).
Then the phase shift for $\pi \pi $ scattering  in the centre of mass is
 $\tan\delta=-L^2 E \gamma\epsilon/48$ to leading order in  $L^{-1}$
which yields a central value $\tan\delta=-0.109$ (including the known
higher order  corrections~\cite{rumm} in $L^{-1}$ gives
$\tan\delta=-0.131$).
 Now, for a nearby particle pole at $E=m_{\rho}$, one can describe the
phase shift  $\delta$ by using the expression for elastic $\pi \pi$
scattering dominated by  this pole: 
 \be
 \tan \delta = { \Gamma(k) \over 2 (m_{\rho}-E)}
 \label{eq.pole}
 \ee
  where the phase shift is  negative because the pole is below the two
body energy.  Here $\Gamma(k)$ is  the decay width parametrised as
$\bar{g}^2 k^3/(E m_{\rho})$ and evaluated with decay momentum $k$ and
energy $E$. Using $a m_{\rho}=0.508$, $Ea=0.709$ and $ka=0.198$, this
gives a coupling of $\bar{g}=1.56^{+21}_{-13}$ which is similar to the
value from our simple analysis above. Indeed in the limit of a weak
transition  and nearly degenerate  energy levels, the two approaches
would give exactly the same result.

Experimental data exist  for the decays $\rho \to \pi \pi$, $K^* \to K
\pi$ and $ \phi \to K \bar{K}$ which all involve the creation of a light
quark pair. This gives us some information on the  dependence of the
coupling strength on quark mass as shown in table~\ref{tab.g} (here the
coupling $\bar{g}$ is normalised to the quark diagrams present in $\rho$
decay). In the limit of a heavy spectator quark  (as for $B^* \to B \pi$
for example), $\bar{g}^2$ is expected to increase  like $m_Q$ which is
the heavy quark mass. This implies that the  coupling does indeed 
increase with spectator quark mass. 
 We also quote in table~\ref{tab.g} the evidence for the  $\rho \to K
\bar{K}$ coupling coming from Regge analyses~\cite{regger} which  bears
on the spectator quark dependence. It will be interesting to use lattice
studies  to explore this further.

\begin{table}
 \begin{center}
 \begin{tabular}{cccl}
 \hline
 $\bar{g}$ & $q({\rm sea})$ & $q({\rm val})$ & method \\
 \hline
 $1.40^{+47}_{-23}$ & s & s & eqs.\ref{eq.xt},\ref{eq.width} \\
 $1.56^{+21}_{-13}$ & s & s &  $\epsilon(\rho)$, eq.\ref{eq.pole} \\
 \hline
 $1.39            $ & n & n & $\rho \to \pi \pi$\\
 $1.44            $ & n & n/s & $K^* \to K \pi$\\
 $1.52            $ & n & s & $\phi \to \bar{K} K$\\
 $1.46-1.74       $ & s & n & $\rho \bar{K} K$ Regge\\
\hline
 \end{tabular}
 \caption{The vector-pseudoscalar-pseudoscalar coupling $\bar{g}$
  }
 \label{tab.g}
 \end{center}
 \end{table}

\section{Conclusions}

 The lattice measurement passes all cross-checks: three different
estimates  of the transition amplitude $x$ are presented, of which the
three-point analysis of $\rho_1 \to \pi_1 \pi_0$ is the most
comprehensive, yielding a coupling  $\bar{g}=1.40^{+47}_{-23}$. The
energy shift of the  $\rho$ state with momentum $2\pi/L$ for different
polarisation directions is seen directly which is a powerful 
cross-check and yields  a coupling $\bar{g}=1.60^{+21}_{-13}$. These
values are close to the  experimental value (for light quark pair
creation) of 1.5.

 Our result is in a finite volume and it would be  appropriate to test
it by using a larger volume, consistent with having the same relative
momentum:  this implies doubling the lattice spatial dimension in at
least one direction. Alternatively  it is possible to explore $\rho_0
\to \pi_{1/2} \pi_{1/2}$  using antiperiodic spatial boundary
conditions~\cite{kc} which could give a cross check. These avenues
require new dynamical quark simulations so are a major computational
endeavour.
 A further step needed would be to extract the continuum limit
of this lattice result. This  needs finer lattice spacing  and the same
physical volume: so again substantial resources. We note that we  have
used a NP clover formulation which does remove order $a$ corrections.

 Even though our method can be affected by excited state contamination, 
the many cross checks we have made are convincing evidence that they are
under control. The systematic errors are still very difficult to
estimate  since we have not made a continuum extrapolation or explored
increasing the  lattice volume. Moreover we have not extrapolated in sea
quark mass, which is  possibly the biggest source of systematic error.
Our result is for quark pair production with quarks of mass 
approximately that of strange quarks.

 Since we find a strong transition for $\rho \to \pi \pi$, in the sense 
that $xt$ is big compared to unity, we have to rely on a mixing  model
to estimate most accurately the coupling strength. This would not be the
case for a weaker decay. Even though the mixing is strong, our estimate
of the $\pi \pi$ energy shift is that it is 0.02 in lattice units, which
will be very  difficult to measure accurately.

 We have shown that hadronic transitions can be explored in lattice QCD 
and the result obtained is consistent with phenomenological values. This
 supports previous studies of unknown phenomena (eg. hybrid meson decay)
and will  allow studies of scalar meson decays which will help to
untangle the confusing experimental situation.



\begin{thebibliography}{99} \frenchspacing


\bibitem{cmdecay} C. Michael,  Nucl. Phys. B327, 515 (1989).

 \bibitem{lu91} M. L\"uscher, Commun. Math. Phys. 104 (1986) 177, ibid.
105 (1986) 153;  Nucl. Phys B354 (1991) 531, ibid. B364 (1991) 237. 

\bibitem{wein} J. Sexton,  A.  Vaccarino  and D.  Weingarten,
{ Nucl.\ Phys.\ B (Proc.\ Suppl.)} {42} (1995) 279;
Phys. Rev. Lett. 75 (1995) 4563

\bibitem{degrand} R. D. Loft and T. A. DeGrand, Phys. Rev. D39 (1989) 2692.  

\bibitem{cmcmscalar} C. McNeile and C. Michael, Phys. Rev. D63  (2001)
114503.

\bibitem{hdecay} UKQCD Collaboration,  C. McNeile, C. Michael and P.
Pennanen, Phys. Rev. D65 (2002) 094505.

\bibitem{cmadj}  C. Michael,  Nucl. Phys. (Proc. Suppl.) B6 (1992) 417-9.
		 
\bibitem{cmpp} P. Pennanen and C. Michael,  hep-lat/0001015.

\bibitem{ukqcd} UKQCD Collaboration,  C. R. Allton et al., Phys Rev D65
(2002) 054502; hep-lat/0107021.

\bibitem{ll} L. Lellouch and M. L\"uscher, Commun. Math. Phys. 219,31
(2001); hep-lat/0003023.

\bibitem{testa} C.-J.D. Lin, G. Martinelli, C.T. Sachrajda and M. Testa,
 Nucl. Phys. B619 (2001) 467; hep-lat/0104006.

\bibitem{fm} UKQCD Collaboration,  M. Foster and C. Michael, 
Phys. Rev. D59 (1999) 074503.

\bibitem{rumm} K. Rummikainen and S. Gottlieb, Nucl. Phys. B 450 (1995) 397.

\bibitem{milc} C. Bernard et al., Phys. Rev. D48 (1993) 4419.

\bibitem{kc} C. Kim and N. Christ, hep-lat/0210003.

\bibitem{regger} G. Ebel et al., Nucl. Phys. B33, (1971) 317.

\end{thebibliography}
\end{document}